\documentclass[a4paper,12pt]{article}

\usepackage{amsmath}
\usepackage{amssymb}
\usepackage{amsthm}
\usepackage{graphicx}
\usepackage[OT4]{fontenc}
\usepackage[utf8]{inputenc}
\usepackage{tikz}
\usepackage{url}
\usepackage[numbers]{natbib}
\usepackage{bm}
\usepackage[margin=3.5cm]{geometry}
\allowdisplaybreaks

\newcommand{\mean}[1]{\mathbf{E}\, #1}

\author{{\L}ukasz D\k{e}bowski%
  \thanks{{\L}. D\k{e}bowski is with the
    Institute of Computer Science, Polish Academy of Sciences,
    ul. Jana Kazimierza 5, 01-248 Warszawa, Poland (e-mail:
    ldebowsk@ipipan.waw.pl). %
  } }

\title{Corrections to ``Universal Densities Exist for Every Finite
  Reference Measure''\footnote{The corrected work was supported by the
    National Science Centre Poland grant 2018/31/B/HS1/04018. We tried
    submitting this correction to the IEEE Transactions on Information
    Theory but we were advised by the Editor to post it to
    ArXiv.org instead.}} \date{}

\begin{document}

\begin{titlepage}

\maketitle

\begin{abstract}
  In the article ``Universal Densities Exist for Every Finite
  Reference Measure'' (IEEE Trans.\ Inform.\ Theory, vol.\ 69, no.\ 8,
  pp.\ 5277–5288, 2023) we neglected to mention relevant contributions
  of Boris Ryabko. We cited a source by him that contains a
  construction of the universal density that we claimed to be our own
  invention without checking the source after drafting the
  article. Our article was motivated by a preprint by Feutrill and
  Roughan, about which we had learned when reviewing the PhD thesis by
  Andrew Feutrill. Whereas we were not allowed to contact Feutrill and
  Roughan besides the review form, we developed some ideas of theirs
  further, ignoring that we stepped into the area previously
  researched by Ryabko. Our published results exceed those by Ryabko
  but the article should have been refocused to report Ryabko's
  contributions. In this note, we detail our citation mistakes.
  \\[1ex]
  \\[1ex]
\end{abstract}


\end{titlepage}



In the recent article \cite{Debowski23b} we neglected to mention
relevant contributions of Boris Ryabko. Worse, in the very beginning,
we cited without rereading a book co-authored by him
\cite{RyabkoAstolaMalyutov16} that contains the central example of the
universal density that we deemed to be our own invention.  Article
\cite{Debowski23b} was directly motivated by a recent paper by
Feutrill and Roughan \cite{FeutrillRoughan21}, about which we had
learned when reviewing the PhD thesis by Andrew Feutrill. Whereas we
have been allowed to contact neither Andrew Feutrill nor Matthew
Roughan, his supervisor, except for the review form, we pursued ideas
of Feutrill and Roughan further. Doing so, we ignored that we entered
the area previously explored by Ryabko. Our results published in
\cite{Debowski23b} exceed those by Boris Ryabko but the introduction
of the article should have been largely rewritten to report Ryabko's
contributions. We regret that having constructed a general theory, we
overlooked to check easily available sources to control for its
novelty.

Article \cite{Debowski23b} concerns universal densities which are
built upon universal measures.  Universal measures are probability
measures on semi-infinite sequences that consistently estimate the
entropy rate for stationary ergodic sources over a finite alphabet.
We speak of measures rather than codes. For probability measures we
require the consistency conditions of the Kolmogorov process
theorem. Besides universal measures on semi-infinite sequences such as
the Prediction by Partial Matching (PPM) measure described in
\cite{Ryabko88en2}, we also have universal distributions on finite
strings that do not satisfy the Kolmogorov process theorem conditions,
such as ones by \cite{Ryabko84en2} based on Shtarkov's minimax
distributions, later described in \cite{Shtarkov87en2}. It should be
stressed that universal densities are based on universal measures
rather than distributions or codes, such as the Lempel-Ziv code
\cite{ZivLempel77,ZivLempel78}. The Kolmogorov process theorem
conditions were important for applications of universal densities in
\cite{Debowski23b}.

One of our neglections in \cite{Debowski23b} concerns reporting the
history of the PPM measure, which is a mixture of measures universal
for $k$-th order Markov sources where $k$ ranges over natural
numbers. The idea that such a mixture of measures or distributions
yields a measure or a distribution that is universal for all
stationary ergodic sources was developed by Ryabko
\cite{Ryabko84en2}. As some examples of distributions (on finite
strings) that are universal for $k$-th order Markov sources, Ryabko
\cite{Ryabko84en2} mentioned Shtarkov's minimax distributions. A few
months later, Cleary and Witten \cite{ClearyWitten84} exhibited
measures (on semi-infinite sequences) that are universal for $k$-th
order Markov sources, which were called PPM measures later. To be
precise, work \cite{ClearyWitten84} was focused on a practical
implementation and did not contain mathematical proofs.

Unfortunately, in articles \cite{Debowski23b,DebowskiSteifer22}, we
cited papers \cite{ClearyWitten84,Ryabko88en2} but we omitted paper
\cite{Ryabko84en2} since an explicit formula for the PPM measures and
their universal mixture as we needed them can be found as late as in
\cite[Eqs.\ (9) and (11)]{Ryabko88en2}. Boris Ryabko denoted these
measures by $\rho$ in \cite[Eq.\ (11)]{Ryabko88en2} or $R$ in
\cite[Eq.\ (1.25)]{RyabkoAstolaMalyutov16}. By contrast, in our
related works \cite{Debowski23b,DebowskiSteifer22} we kept on calling
these measures the PPM measures.  We preferred this name as more
distinctive than $\rho$ or $R$. It should be noted, however, that
Cleary and Witten \cite{ClearyWitten84} did not consider the infinite
mixture over Markov orders considered by Ryabko in \cite{Ryabko84en2}
and \cite[Eq.\ (11)]{Ryabko88en2}, which is necessary to guarantee the
universality of the PPM mixture measure.  In our presentations of the
PPM measures in \cite{Debowski23b,DebowskiSteifer22}, we also applied
some modifications of \cite[Eqs.\ (9) and (11)]{Ryabko88en2}: We used
the Laplace (+1) smoothing rather than the Krichevsky-Trofimov (+1/2)
smoothing, the latter being minimax optimal, cf.\ \cite[page
8]{RyabkoAstolaMalyutov16}. Instead of irrational weights
$1/\log(k+1)-1/\log(k+2)$, cf.\ \cite[page
10]{RyabkoAstolaMalyutov16}, we used rational ones $1/(k+1)-1/(k+2)$,
which mattered for computability analyses in \cite{DebowskiSteifer22}.

The more important neglection in \cite{Debowski23b} concerns the
example of a universal density which we call the Non-Parametric
Differential (NPD) density after Feutrill and Roughan
\cite{FeutrillRoughan21}. They proposed combining consistent entropy
estimators for a countable alphabet by Kontoyiannis et
al. \cite{KontoyiannisOthers98} with a quantization of the time series
to obtain a biased estimator of the differential entropy rate. They
called this estimator the NPD estimator. Inspired by
\cite{FeutrillRoughan21} and \cite{Ryabko88en2}, in \cite{Debowski23b}
we proposed the NPD density, being a mixture of quantized estimators
over infinitely many nested quantization levels, which yields a
consistent estimator of the differential entropy rate.

However, this quantization idea had been previously discovered by
Boris Ryabko. The same construction as our NPD density can be found in
\cite[Eqs. (15) and (19)]{Ryabko09}, which we did not cite in
\cite{Debowski23b}, and also is present in \cite[Section
1.6.1]{RyabkoAstolaMalyutov16}, the book that we cited in the
beginning of \cite{Debowski23b} without rereading it after drafting
that article.  As a result, article \cite{Debowski23b} was written
from the wrong perspective that the NPD density was our own extension
of the supposedly original quantization idea by Feutrill and Roughan
\cite{FeutrillRoughan21}, which was analogous to Ryabko's extension
\cite{Ryabko88en2} of the Markov approximation idea by Cleary and
Witten \cite{ClearyWitten84}. Such a claim does an injustice to Boris
Ryabko, who proposed both the PPM measure, being a mixture over Markov
orders \cite{Ryabko84en2,Ryabko88en2}, and the NPD density, being a
mixture over quantization levels \cite{Ryabko09}.

Independent discoveries sometimes lead to stronger results, however.
It seems that this is the case of our paper \cite{Debowski23b}. Ryabko
\cite[Theorem 2 and Claim 3]{Ryabko09}, see also \cite[Section
1.6.1]{RyabkoAstolaMalyutov16}, only showed that the NPD density is
universal in the class of stationary ergodic Markov time series. In
\cite[Theorem 5]{Debowski23b}, we managed to demonstrate that the NPD
density is universal in the whole class of stationary ergodic time
series. Our proof rests on the monotone convergence of $f$-divergences
for filtrations \cite[Lemma 1]{Debowski23b}, \cite[Chapter 3, Problem
4]{Debowski21}, cf. also \cite[Lemma 2]{Debowski09}, which is likely a
known fact.

Having proved universality of the NPD density, in \cite{Debowski23b}
we considered various applications thereof to marginal density
estimation, to universal prediction with the $0-1$ loss (strengthening
our earlier results of \cite{DebowskiSteifer22}), as well as special
cases of consistent entropy estimators for some processes over a
countable alphabet and for Gaussian processes. Whereas these results
exceed Boris Ryabko's results described in
\cite{Ryabko09,RyabkoAstolaMalyutov16}, it is important to stress that
all our results pertain only to stationary ergodic processes with a
finite differential entropy rate with respect to the chosen finite
reference measure.

In this context, we should have recalled this counterexample: Let
$(\mathbb{X},\mathcal{X})$ be a finite measurable space. Consider the
product space $(\mathbb{X}^{\mathbb{Z}},\mathcal{X}^{\mathbb{Z}})$ and
put random variables
$X_k:\mathbb{X}^{\mathbb{Z}}\ni(x_i)_{i\in\mathbb{Z}}\mapsto
x_k\in\mathbb{X}$.  We write strings $x_{j:k}:=(x_j,x_{j+1},...,x_k)$.
For any probability measure $\bar R$ on
$(\mathbb{X}^{\mathbb{Z}},\mathcal{X}^{\mathbb{Z}})$ there exists a
stationary ergodic measure $P$ on
$(\mathbb{X}^{\mathbb{Z}},\mathcal{X}^{\mathbb{Z}})$ such that
\begin{align}
  \label{NoPredictor}
  \limsup_{n\to\infty}
  \log\frac{P(X_{n+1}|X_{1:n})}{\bar R(X_{n+1}|X_{1:n})}>0
\end{align}
with a positive $P$-probability \cite[Proposition 3]{Ryabko88en2}.
By contrast, in \cite[Definition 6 and Theorems 6 and 7]{Debowski23b},
applying the idea of Gy\"orfi et al.\ \cite{GyorfiPaliMeulen94}, we
constructed the Ces\`aro mean measure $\bar R$ such that
\begin{align}
  \label{Predictor}
  \lim_{n\to\infty}
    \mean\log\frac{P(X_{n+1}|X_{1:n})}{\bar R(X_{n+1}|X_{1:n})}=0
\end{align}
for any stationary ergodic $P$.  Claims (\ref{NoPredictor}) and
(\ref{Predictor}) may seem contradictory but they are not if the
oscillations of sequence
$\log P(X_{n+1}|X_{1:n})-\log \bar R(X_{n+1}|X_{1:n})$ become more and
more sparse. As we showed in \cite[Theorem 8]{Debowski23b}, the
Ces\`aro mean measure $\bar R$ can be constructed from any universal
density and it induces a universal predictor with the $0-1$ loss. The
total number of mistakes of this predictor may be unbounded but their
density approaches zero.

In passing, we notice that in Definitions 7--9 of \cite{Debowski23b},
which define various universal predictors, we pasted thrice the same
term ``strongly universal'' instead of ``strongly universal'',
``universal in expectation'', and ``universal in probability'',
respectively. These definitions should be analogous to Definitions
1--3 of \cite{Debowski23b}, which define universal measures.

\section*{Acknowledgments}

I thank Boris Ryabko for recalling me his works after the publication
of \cite{Debowski23b}. 

\bibliographystyle{IEEEtran}

\bibliography{0-journals-abbrv,0-publishers-abbrv,ai,ql,math,mine,tcs,books}

\end{document}